\documentclass[twocolumn,aps,prl,showpacs]{revtex4}
\usepackage{graphicx}
\usepackage{dcolumn}
\usepackage{latexsym}
\begin{document}

\title{
Optimization by thermal cycling
}
\author{A.~M\"obius$^1$, A.~Neklioudov$^1$, 
A.~D\'{\i}az-S\'anchez$^{1,2}$, K.H.~Hoffmann$^3$, A.~Fachat$^3$, and 
M.~Schreiber$^3$}
\address{
$^1$Institut f\"ur Festk\"orper- und Werkstofforschung,
D-01171 Dresden, Germany,\\
$^2$Departamento de F\'{\i}sica, Universidad de Murcia, E-30071 
Murcia, Spain,\\
$^3$Institut f\"ur Physik, Technische Universit\"at Chemnitz,
D-09107 Chemnitz, Germany
}
\date{August 15, 1997}
\begin{abstract}
An optimization algorithm is presented which consists of cyclically
heating and quenching by Metropolis and local search procedures, 
respectively. It works particularly well when it is applied to an 
archive of samples instead of to a single one. We demonstrate for the 
traveling salesman problem that this algorithm is far more efficient 
than usual simulated annealing; our implementation can compete 
concerning speed with recent, very fast genetic local search 
algorithms. 
\end{abstract}


\maketitle
\narrowtext

Combinatorial optimization problems have been a challenge for a long 
time. One of the most popular is the traveling salesman problem (TSP): 
how to find the shortest roundtrip through a given set of cities. It 
serves as a standard test for new optimization algorithms, for surveys 
see \onlinecite{John.McGe,Jung.etal}. Many combinatorial optimization 
tasks are of considerable practical importance. Thus algorithms are 
highly desirable which yield good approximations of the exact solution 
within a reasonable CPU time, and which need only a modest effort in 
programming. As such, simulated annealing \cite{Kirk.etal,Laar.Aart} 
has become a standard procedure. Genetic algorithms have been 
increasingly successful, in particular such algorithms where local
minima are the species considered, see 
\onlinecite{Muhl.etal,Frei.Merz,Merz.Frei} and references therein. 
Furthermore, e.g.\ alternately optimizing the system itself and 
various subsystems of it proved to be effective \cite{Ditt}. Still, 
the danger of getting trapped in any local instead of the global 
minimum increases with the size of the problem and cannot be overcome 
by these heuristic methods.

Simultaneously, exact methods have been developed further mainly using 
branch-and-bound and branch-and-cut ideas \cite{Bala.Toth,Jung.etal2}.
Even a specific TSP instance including 7397 cities could be solved 
\cite{John.McGe,Rein}. However, for a given size, the effort 
necessary to find the exact solution can vary enormously from problem 
to problem; the TSPLIB95 \cite{Rein} includes an instance of 1577 
cities which has not been solved exactly yet. The ideas basic in exact 
algorithms can also be used as background for fast local search 
procedures. Applied to randomly chosen states, such restricted 
branch-and-bound search implementations can compete with simulated 
annealing with optimized schedule \cite{Moeb.Rich}. 

Here, we present an algorithm which combines the advantages of 
simulated annealing with those of fast local search procedures. 
Additionally, our algorithm utilizes an archive of local minima 
similarly to a recently published hybrid-Metropolis procedure 
\cite{Moeb.Thom}. First, we explain the general ideas of this 
algorithm. Then we illustrate its efficiency by considering several 
TSP instances. 

Simulated annealing \cite{Kirk.etal} can be understood as a random 
journey of the sample (i.e.\ the approximate solution) through a 
hilly landscape formed by the states of its configuration space. The 
altitude, in the sense of a potential energy, corresponds to the 
quantity to be optimized. In the course of the journey, the altitude 
region accessible with a certain probability within a given number of 
steps shrinks gradually due to the decrease of the temperature in the 
Metropolis simulation \cite{Metr.Rose} involved. The accessible area,
i.e., the correponding configuration space volume, thus shrinks until 
the sample gets trapped in one of the local minima. 

Deep valleys attract the sample mainly by their area. However, it is 
tempting to make use of their depth. For that, we substitute the 
slow cooling down by a cyclic process: First, starting from the lowest 
state obtained so far, we randomly deposit energy into the sample by 
means of a Metropolis process with a certain temperature $T$, which is 
terminated, however, after a small number of steps. This part is 
referred to as heating. Then we quench the sample by means of a local 
search algorithm. This is an iterative process which consists in 
searching systematically the neighborhood (defined via the move class) 
of the actual state for states of lower energy, where always the first 
such state found is accepted. It stops when no lower neighboring state 
exists. Heating and quenching are cyclically repeated; the amount of 
energy (controlled by $T$) deposited in a cycle decreases gradually. 
This process continues until, within a `reasonable' CPU time, no 
further improvement can be found. 

Our procedure has some resemblance to genetic local search algorithms 
\cite{Muhl.etal,Frei.Merz,Merz.Frei} as well as to the iterated 
Lin-Kernighan concept \cite{John.McGe,Mart.etal,Mart.Otto}. But there 
are  substantial differences. In genetic local search algorithms, 
crossover is the process to produce long jumps overcoming barriers in 
the configuration space. In the iterated Lin-Kernighan approach, the 
sample is kicked out of the local minimum by moves chosen from a 
specific class not considered in local optimization. Our algorithm, 
however, is not based on introducing a class of particular elementary 
moves for crossing barriers. Instead, it makes use of an inherent 
feature of combinatorial optimization problems: whether or not a 
certain move is permitted, or reduces energy, depends on the actual 
state. Loosely speaking, the moves do not commute. Thus, provided $T$ 
is large enough, even exciting the sample by a sequence of several 
rather primitive moves of a type used also in quenching should be 
sufficient for the cycle returning to its initial state only very 
seldom. This simple concept should be applicable to all problems which 
can be treated by simulated annealing. Moreover, this approach permits 
to make good use of certain complex moves, which would diminish the 
efficiency of heating due to their huge number, by considering them 
only in quenching. 

When our algorithm is applied to a specific problem, several choices
have to be made. The first one concerns the starting temperature, 
cf.\ \onlinecite{Mari}. One can relate it to the energy difference 
between random and quenched states, or to the local misfit in quenched 
states, in the sense of frustration of a spin glass.

The next issue is the amount of energy deposited into the sample while 
heating. Two contradicting demands have to be met: the gains of the 
previous cycles should be retained, but the modifications must be 
sufficiently large, so that another valley can be reached. Thus the 
heating process has to be terminated in an early stage of the 
equilibration. An effective method is to stop it after a fixed number 
of successful Metropolis steps. 

Our recipe for controlling $T$ makes use of the rate of finding final 
states of the heating-quenching cycles which have lower energies than 
all previous ones: we keep $T$ constant as long as this rate exceeds a 
certain value. The lower this value, the better should on average be 
the final approximate solution.

The number of modifications performed within one heating process
is small; moreover, only few of the cycles yield an improvement of the 
best state so far. Thus a considerable accelaration can be achieved by 
a reduction of the move class considered in heating: after each change 
of $T$, we tabulate those moves which excite the best state so far by 
energies not larger than $T$. While heating, we consider only these 
moves. However, during quenching, we consider the related complete 
move class.

Finally, one has to determine a stopping criterion. The most natural 
way is to compare the probability that the cycle returns the sample 
to its initial state (or to one with the same energy) with the 
probability of finding states of lower energy. 

We improve now the above approach by searching the configuration space
starting the cycles at random from one of $N_{\rm a}$ local minima 
held in an archive \cite{Moeb.Thom}, instead of from the best state so 
far. In this simultaneous search, the above described table of moves 
for heating comprises the exciations of all archive states by energies 
up to $T$. We reduce this table utilizing `collective experience': If 
all archive states have a certain common feature, the global minimum 
is likely to also have this feature. Therefore, after changing $T$, we 
determine those degrees of freedom which are frozen out, and eliminate 
all excitations from the table which concern one of those. This 
reduction corresponds to the use of `don't look bits' 
\cite{Merz.Frei}, or to the search for backbones \cite{Schn.etal}. 

In such a simultaneous search, it is tempting to simulate
evolution by mutation and competition: If the energy of the final 
state of a cycle is smaller than that of the highest archive state, 
one would substitute this final state for the highest archive state. 
However, we observed that it is mostly preferable to replace the 
initial state of the actual cycle instead of the highest archive 
state. 

Concluding the general part we mention two other interpretations of 
such an algorithm. In the biological context, our procedure is of the 
genetic local search type \cite{Muhl.etal,Frei.Merz,Merz.Frei}. But it 
is based on mutations only, crossovers are not used. From an 
economist's viewpoint \cite{Hube.etal}, it increases the performance 
by means of risk reduction on two levels: (i) After leaving a local 
minimum, we perform only a small number of modifications before we 
decide whether or not the continuation of this way is promising. 
(ii) Several valleys are investigated simultaneously. 

We have applied the above ideas to a series of symmetric traveling 
salesman problems, as well as to the Coulomb glass. In both cases, the 
corresponding codes work very efficiently. Here, we focus onto the 
TSP, where the length of the roundtrip has the meaning of the energy 
to be minimized. We use the following notions: tour and subtour denote 
closed roundtrips through all cities or part of the cities, 
respectively, whereas chains and subchains stand for tours and 
subtours with one connection eliminated, respectively.

The efficiency of our procedure, as well as that of simulated 
annealing, depends to a large extent on the move class taken into 
account. We consider two elementary heating processes: (i) cutting out 
from the tour a subchain, reversing its direction, and inserting it 
again at the same place; (ii) shifting a city from one to another 
position in the tour. However, we do not consider all such moves, but 
use the above introduced restrictions.

In quenching, we choose between four possibilities concerning the
kind of metastability to be reached:\\ 
(a) stable with respect to reverse of a subchain, as well as 
to a shift of a city (unrestricted move class of heating);\\
(b) same as (a), and stable with respect to cutting three connections 
of the tour, and concatenating the three subchains in a new manner;\\
(c) same as (b), and stable with respect to first cutting the tour 
twice and forming two separated subtours, and connecting then these 
subtours after cutting two other connections;\\
(d) same as (c), and stable concerning a restricted
Lin-Kernighan search \cite{Lin.Kern} which consists in cutting the 
tour once, then several times alternately cutting the chain and 
concatenating the subchains, and finally connecting the ends of the 
chain again, where the number of trials to modify the chain is 
restricted to 1000.\\
The efficiency of our local search program rests on trying new 
connections according to increasing length, where appropriate bounds 
are utilized to terminate the search as soon as it becomes useless. 
Moreover, for stage (c), we demand that the sum of the lengths of the 
two closed subtours is shorter than the length of the original tour.

Our algorithm includes several adjustable parameters. They all 
influence simultaneously both the quality of the final result and the
CPU time needed. Nevertheless, we have observed that they affect only 
rather weakly the performance of the algorithm, i.e., the quality of 
the averaged final result obtained for the archive size $N_{\rm a}$ 
adjusted so that the program finishes within a certain CPU time. For 
the TSP instances considered, the following values were good choices. 
First, we quench $50 N_{\rm a}$ times a randomly chosen tour, and set 
up the initial archive from the shortest $N_{\rm a}$ of the tours 
obtained. As starting temperature, we choose the reduction of the 
length when quenching a random tour, divided by the number of cities. 
In heating, we modify the tour always 50 times. At each $T$, we 
perform $5 N_{\rm a}$ heating-quenching cycles. If, during these 
cycles, an archive state has been replaced, we perform again 
$5 N_{\rm a}$ cycles, and so on. Otherwise, $T$ is reduced by a factor 
of 0.9. We terminate the process when it has returned $10 N_{\rm a}$ 
times to an archive state (or to a state with equal energy) without 
any substitution of an archive state. 

We tested our program considering five symmetric TSP instances out of 
the TSPLIB95 \cite{Rein}, including between 442 and 3795 cities. 
We performed several tests with different numbers of archive states,
quenching always to metastablity (d) defined above. The results are 
given in Table I. They demonstrate the high efficiency of our 
algorithm: For the instances with 442, 532, and 783 cities, the known 
exact optima were reproduced. The instances with 1577 and 3795 cities
have not been solved exactly yet. Here, our best tour lengths agree 
with the known upper bounds for the optimum tour length \cite{Rein}. 

\begin{table}

\begin{tabular}{rrrrrr}
Problem&$N_{\rm a}$&$L_{\rm {min}}$&$L_{\rm {max}}$&$L_{\rm {mean}}$&
$\tau_{\rm {CPU}}$\\ 
\hline
pcb442&1&50778&51004&50849&25\\
pcb442&3&50778&50912&50794&69\\
pcb442&5&50778&50778&50778&145\\ \hline
att532&1&27686&27778&27722&32\\
att532&3&27686&27732&27707&88\\
att532&5&27686&27715&27694&191\\
att532&8&27686&27705&27692&372\\ 
att532&12&27686&27693&27686.4&645\\ \hline
rat783&1&8809&8829&8818&38\\
rat783&3&8806&8823&8812&101\\
rat783&5&8806&8810&8806.7&213\\
rat783&8&8806&8809&8806.1&427\\ 
rat783&12&8806&8806&8806&759\\ \hline
fl1577&1&22254&22289&22264&251\\
fl1577&3&22250&22267&22256&808\\
fl1577&5&22249&22261&22252.5&1730\\
fl1577&8&22249&22255&22251.8&3350\\ 
fl1577&12&22249&22253&22250.4&6070\\ \hline
fl3795&1&28775&28890&28809&2100\\ 
fl3795&3&28775&28830&28790&6300\\ 
fl3795&5&28772&28784&28776.5&13500\\ 
fl3795&8&28772&28780&28774.8&25900\\ 
\end{tabular}

\caption{
Dependence of the tour length of the approximate solution, and of the 
CPU time on the archive size $N_{\rm a}$ for five instances out of the 
TSPLIB95 \protect\cite{Rein}. Smallest, largest, and mean tour 
lengths, $L_{\rm {min}}$, $L_{\rm {max}}$, and $L_{\rm {mean}}$, are 
given for series of 20 runs. $\tau_{\rm {CPU}}$ denotes the CPU time 
in seconds for a single run using one PA8000 180 MHz processor of an 
HP K460. The local search  guaranteed stability with respect to
demand (d) defined in the text.
}
\end{table}

However, our algorithm yields generally only an approximate solution. 
Thus the truly important property is the relation between mean 
quality of the solution and the necessary CPU time. We compared the
performance of our program with that of the genetic local search 
algorithm \cite{Merz.Frei} by Merz and Freisleben - a significantly 
improved version of their winning algorithm of the {\it First 
International Contest on Evolutionary Optimization} \cite{Bers.etal} 
-, and with the performance of the iterated Lin-Kernighan code by 
Johnson and McGeoch, illustrated by Table 16 of 
\onlinecite{John.McGe}. (Our CPU is by factors of roughly 1.7 and
3 faster than the CPUs used in \onlinecite{Merz.Frei} and 
\onlinecite{John.McGe}, respectively.)\\
- Compared to \onlinecite{Merz.Frei}, our code has roughly the same 
efficiency for the three instances with several hundred cities. 
However, for the instances with 1577 and 3795 cities, our program is 
more efficient. In particular, for the task fl3795, it yields a 
significantly better mean value in less than one fourth of the CPU 
time.\\
- Compared to \onlinecite{John.McGe}, our program is by a factor 5 to 
10 slower for pcb442 and att532 if the accuracy of our results for 
$N_a = 3$ is achieved. The performance gap seems to shrink with 
increasing accuracy demand \cite{John}. However, for fl3795, our code 
performs considerably better.\\
The clear advantage of our program for fl3795 in both comparisons can 
arise from particular robustness, or from good scaling properties. 
Further investigations are needed to clarify this point, and to decide 
whether the advantage is caused by the use of thermal cycling to 
overcome barriers, or by a feature of our local search procedure. 

One should also make comparison with the state of the art exact
algorithms. For the Padberg-Rinaldi 532 city problem, the 
branch-and-cut program by Thienel and Naddef, one of the presently 
fastest exact codes, needs 16.5 minutes at a SPARC10 \cite{Thie}, 
corresponding to roughly 4 minutes for our CPU. Utilizing an archive of 
12 states and cyclically quenching to stage (d) defined above, we 
performed 100 runs. Our Monte Carlo approach reproduced the optimum tour 
length 27686 in 95 of these runs, where it needed 11 CPU minutes for one 
of them. In the residual five cases, we obtained tours with length 
27693. 

\begin{figure}
\includegraphics[width=0.85\linewidth]{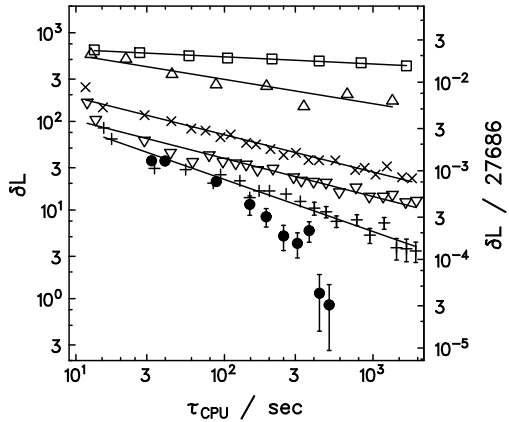}
\caption{
Relation between CPU time, $\tau_{\rm {CPU}}$ (in seconds for one
PA8000 180 MHz processor of an HP K460), and deviation, 
$\delta L = L_{\rm {mean}} - 27686$, of the obtained mean approximate 
solution from the optimum tour length for the Padberg-Rinaldi 532 city 
problem. $\Box$: repeated quench to stability with respect to demand 
(a) defined in the text; $\bigtriangleup$: simulated annealing; 
$\times$, $\bigtriangledown$, $+$, $\bullet$: thermal cycling with 
archives of various size, and local search concerning conditions (a), 
(b), (c), and (d), respectively. In all cases, averages were taken 
from 20 runs. Errors ($1\sigma$-region) are presented if they exceed 
the symbol size. The straight lines are guides to the eye only.
}
\label{fig1}
\end{figure}

We have shown above that thermal cycling can be used as basis for 
state of the art Monte Carlo optimizations. Now, we demonstrate that 
this physical approach can be far more effective than standard 
concepts, even if only a comparatively small amount of time is 
invested in programming. For the Padberg-Rinaldi 532 city problem, 
Fig.\ 1 confronts four versions of our algorithm, differing from each 
other by the complexity of the local search, with usual simulated 
annealing, and with repeatedly quenching randomly chosen states. 
(Simulated annealing uses the same moves as heating above, but trials 
were restricted to the 30 nearest cities.) The figure shows that the 
thermal cycling procedure proposed is far faster than usual simulated 
annealing, even if the local search leads only to the `primitive' 
metastablity (a).

Concluding, we have shown that simulation of thermal cycling is an
effective tool for treating combinatorial optimization problems. Its 
efficiency in finding very good approximations of the optimal solution 
was illustrated by considering the traveling salesman problem. Our 
algorithm, which is well appropriate for parallelising, should also be 
useful for a broad class of other combinatorial optimization problems.

\vspace{.2cm}

This work was supported by the SMWK and DFG (SFB 393). We are 
particularly indebted to B.~Freisleben and S.~Thienel for valuable 
literature hints, as well as to M.~Pollak and U.~R\"o{\ss}ler for a 
series of critical remarks. Moreover, discussions with F.-M.~Dittes, 
M.~Golden, D.S.~Johnson, S.~Kobe, A.~Hartwig, M.~Ortu\~no, M.~Richter, 
and J.~Talamantes were very useful.

\end{document}